\documentclass[runningheads]{llncs}
\usepackage{graphicx}
\usepackage{todonotes}
\usepackage{siunitx}
\usepackage{graphicx}

\begin{document}
\title{Exploration of Voice User Interfaces for Older Adults - A Pilot Study to Address Progressive Vision Loss}
\titlerunning{Voice Assistants and People with Progressive Vision Loss}
%

\author{Anna Jaskulska\inst{1} \orcidID{0000-0002-2539-3934}\and
Kinga Skorupska\inst{2}\orcidID{0000-0002-9005-0348}\and
Barbara Karpowicz\inst{2}\orcidID{0000-0002-7478-7374}\and
Cezary Biele\inst{3}\orcidID{0000-0003-4658-5510}\and
Jaros{\l}aw Kowalski\inst{3}\orcidID{0000-0002-1127-2832}\and
Wies\l{}aw~Kope\'{c}\inst{2,}\orcidID{0000-0001-9132-4171}}
\authorrunning{Jaskulska et al.}

\institute{Kobo Association, Warsaw, Poland\\
\email{a.jaskulska@kobo.org.pl}\\
\and
Polish-Japanese Academy of Information Technology, Warsaw, Poland
\and
National Information Processing Institute, Warsaw, Poland\\
}

\maketitle              
\begin{abstract}
Voice User Interfaces (VUIs) owing to recent developments in Artificial Intelligence (AI) and Natural Language Processing (NLP), are becoming increasingly intuitive and functional. They are especially promising for older adults, also with special needs, as VUIs remove some barriers related to access to Information and Communications Technology (ICT) solutions. In this pilot study we examine interdisciplinary opportunities in the area of VUIs as assistive technologies, based on an exploratory study with older adults, and a follow-up in-depth pilot study with two participants regarding the needs of people who are gradually losing their sight at a later age.

\end{abstract}

\section{Introduction}

The use of voice interfaces in general is an increasingly popular research topic among many age groups, children \cite{bielevakids2019}, teenagers \cite{vateens2017} and younger and older adults \cite{OA13VAinteraction} \cite{vaforaginginplace2019}. 
One of the key benefits is their accessibility via voice alone, especially as the recent developments in AI and NLP make voice commands more intuitive.
However, most of the research is done with either healthy older adults \cite{portet12VAhealthyoa}, or those who have defined and unchanging impairments. 
During an exploratory study with a group of older adults we have discovered the research niche and, at the same time, a real need to address the problem assisting older adults with progressive vision loss whose degree of impairment is shifting, for whom VUIs may be the assistive technology of choice, because they allow for near-natural interaction, with the use of familiar patterns of speech.

For this reason we decided to explore the potential of voice assistants for people in the process of losing their sight at a later age. These users are at a disadvantage, as on one hand they do not possess the tools and skills of the blind, but on the other conventional technology poses an increasingly greater challenge to them. We believe that this area of research presents multiple interdisciplinary challenges, ranging from the best approaches to training new IVA users, through the psychological challenges associated with some loss of function, to the possibility of introduction of novel ICT solutions easing this transition. In this paper we present a few preliminary insights from an exploratory study with a group of older adults, and a follow-up in-depth interview-based study of the potential of voice interfaces for older adults coping with progressing vision loss.

\section{Literature review}

Voice interaction has been implemented in mobile phones and tablets for several years now, but only recently standalone devices (sometimes called smart-speakers) like Google Home, Amazon Alexa or Apple HomePod, became available. Increasing adoption of Voice User Interface (VUI) and Voice Assistants (VA) and other technology trends like Internet of Things (IoT), Smart Homes and robotics in the context of demographic challenges (aging societies) and digital market opportunities (silver economy) raise numerous questions. 

For example, one of the most recent studies conducted in 2017 in United Kingdom by EngineeringUK \footnote{www.engineeringuk.com] for Tomorrow’s Engineers 2017: (2018)} shows that VA technology proliferation is more than two and a half greater than robotics \ref{tab:table1}. Based on the numerous previous studies on designing interfaces for older adults, e.g. \cite{fiskdesigningolderr,kumar2012typing} followed by the most recent detailed exploration in this field of study \cite{brewer2016would,weber2016what,munteanu2018pronunciation} alongside with depicted tendencies of mainstream voice-based interfaces\cite{pradhan2018accident}, based on numerous research and studies on direct intergenerational engaging of older adults into technology and participatory design \cite{kopec2017living,kopec2017older,kopec2017spiral,skorupska2018smarttv} we decided to explore the area of older adults interactions with voice assistants. Especially that a recent study by Pradhan et al.\cite{pradhan2018accident} demonstrated that voice assistant technology has a great potential for becoming a useful tool for multiple groups of users with varying needs, as voice assistants are already used by people with disabilities. Voice assistants have also been considered for the blind, in a publication by the Department of Ophthalmology, Royal Gwent Hospital, Newport, UK \footnote{The paper is available at Nature.com: https://www.nature.com/articles/eye2017165} where the authors postulate that VA's may be useful for the visually impaired older adults, especially as they are becoming more tech-savvy. VUIs as assistive technolgy are especially potent when paired with Smart homes, given the proliferation of VA and robots as seen in Table \ref{tab:table1} on the example of people in the UK.

\begin{table}
    \centering
    \caption{Proliferation of VA and robots in UK (percent of respondents ever used the technology at home; according to EngineeringUK 2017 study)}
    \label{tab:table1}
    \begin{tabular}{c | c | c}
      & {\small \textbf{robots}}
      & {\small \textbf{VA}} \\
      18-24 & 15\% & 52\% \\
      25-34 & 18\% & 39\% \\
      35-44 & 11\% & 34\% \\
      45-54 & 8\% & 27\% \\
      55-64 & 2\% & 29\% \\
      65+ & 0\% & 0\% \\
      Overall & 14\% & 38\% \\
    \end{tabular}
\end{table}


A literature review on Smart Homes delivered by Tom Hargraves and Richard Huxwell-Baldwin\cite{hargreaves2013uses} shows that there are three key perspectives of looking at IVAs. The first is the functional view which perceives this technology as a way to better manage everyday. The second is instrumental: SHT is a way to reduce energy demand. The third is a socio-technical perspective, where SHT is the next step in the development of a way to full digitalization and electrification of the human environment. The social and technical perspectives perceive SHT not only as a tool for achieving goals and improving the lives of people but also as a next step in the process of changing (evolving) the human environment. Social and individual day-to-day practices can be transformed, including the role of the family and social and professional relationships.

In this paper we follow this reasoning with a glimpse how the older adults wish to use this technology and what are the barriers to its use, the aspirations of this group of users, as well as what opportunities it presents to address the needs of older people with progressive vision loss.

\section{Initial exploratory study}

\subsection{Methods}
The exploratory research involved a semi-structured mixed method scenario of group interviews with direct participant involvement, which is reported at length by Kowalski et al \cite{Kowalski:2019:OAV:3290607.3312973}. The scenario consisted of an introductory session with voice user interfaces, presentation of Google Home (in a setup connected with a smartphone and peripherals like a TV set, an STB, light and so on) and free use of Google Home by the seven participants, whose mean age was 73.14.

There was a discussion during which the participants shared their opinions on this technology and its applications for them and in their homes.

\subsection{Results and Discussion}

Overall, the older adults who took part in our workshops mentioned multiple possible modes of use of the Voice Assistants, such as: timer / alarm / alarm clock, question about the weather, question about information (e.g. about traffic on the road, information from Wikipedia, recipes), turning on the TV set, keeping a shopping list, listening to audiobooks (P1), help in solving crosswords (P4), finding a recipe and cooking assistance (P4), language learning assistance (P6), telling a jokes (P7). When this list is compared, with research done on how Voice Assistants are being used by early adapters in a study by EngineeringUK 2017, one can see a massive overlap, as people often request their help with music (31\%), search (26\%), smart home (15\%), reminders/lists (9\%), shopping (6\%), making calls (5\%),  playing games(3\%), banking (2\%).

As reported in Kowalski et al. \cite{Kowalski:2019:OAV:3290607.3312973} participating older adults saw great potential in voice interfaces and during the brainstorming session could identify areas of use, which already exist in voice assistant programming.

Our older adults were hopeful about how much can be achieved thanks to voice assistants with little training. Unlike modern input devices and graphical user interfaces which rely on precise motor movements and keen eyesight (mouse, small keys and touchscreens with multitouch), these interfaces make use of people's natural ability to converse. On top of the mentioned uses, our group of older adults were also interested in automatizing some tasks that they deemed troublesome (including routine things, such as doing the laundry and washing the dishes) and using support in various areas. 

The participants also pointed to the benefits of using VUI in connection with Smart Home devices for people with various impairments, including difficulties in moving around. They stressed the fact that people requiring a lot of assistance could become partially independent thanks to such solutions. This insight led to the follow-up study described below.

\section{Follow-up in-depth pilot study}

\subsection{Methods}

Having in mind that most of the research is done with either healthy older adults or those with motor disabilities, based on the insights from our aforementioned preliminary studies an additional research area emerged: we decided to explore the potential of voice assistants for older adults in the process of losing their vision. People who gradually lose their sight in old age do not have such abilities as long-term blind people, including those born blind. Moreover, at the same time these older adults are facing many challenges related to ageing itself in such areas as emotional, social, cognitive and physical functioning and "major subgroups of older adults with vision impairment also are in need of high-quality professional support to improve their psychosocial adaptation" \cite{lee_psychological_2013}.
For them it is very difficult to use the previously familiar devices, while one the other hand they lack the training with refreshable braille displays (braille terminals) or screen readers dedicated to the blind. These users have a clear disadvantage, as on one hand they do not possess the tools and skills of the blind, but on the other hand conventional technology poses an increasingly greater challenge to them, however they wish to continue using it. 

\subsubsection{Scenario}
The follow-up consisted of an in-depth exploratory interview - the issues explored in it were inspired by the results of the previous study regarding the topics of interest in relation to voice assistants \cite{Kowalski:2019:OAV:3290607.3312973}. It allowed for flexibility and the interviewer responded to the participants' need to elaborate on or include topics of key importance to them. Of the two invited participants, each had their own experience with progressing vision loss, as indicated below.

\subsubsection{Participants}

To kick start our exploratory study we invited two participants, one of them (P2) took part in our previous research on IVAs \cite{Kowalski:2019:OAV:3290607.3312973}, 
whereas P8 is a new participant, invited by P2. Both of them are older adults with an experience of progressing vision loss, but actively using ICT solutions. P8's vision has been gradually worsening over the years, and now their vision is blurry and they see "as if in a mist", with also narrowing field of vision.
P2 on the other hand, does not have first-hand experience of vision loss, but has the experience of a caretaker, being a close friend of P8 who lives alone, but whose family also often visits. P8's knowledge of English is communicative, as they were required to read professional literature in English for their job, although they have since retired.

\subsection{Results and Discussion}

\subsubsection{On Using the Computer}
Currently P8 has two main problems when using a laptop. First of all it is very difficult for them to find the mouse cursor. Second, they have difficulties using the laptop keyboard, especially with keyboard shortcuts requiring the users to hold a few keys pressed at the same time (which P8 expected to be easier, as they knew touch-typing on a typewriter).
They have the habit of employing the computer for a variety of things, for example, using the optical drive of a computer to play music CDs and Audiobooks.
P8 used to use the built-in magnifying glass program in their computer, and how, to use a smartphone they use a physical magnifying glass. However, they comment: "It's the worst when the magnifying glass is lost" which happens as they get left on the table, because of their narrow field of vision - for this reason they carry a few of them on them at all times. This is why P8 later comments, "one needs to have their head on hinges" meaning, that they have to turn their head a lot to scan their environment.

\subsubsection{On Smart Home Applications}
The aspect of typical Smart Home technology was not crucial for P8, as they have their home well-memorized and they can both remember the location of light switches and see enough to find them. Although they liked the idea of voice controlling their window blinds, this was mentioned as a luxury. 
The TV set was another device P8 wished to control using their voice. However, the same was true of other older adults without significant vision loss and P8 earlier remarked that the TV remote does not pose them troubles.
However, one need that P8 underlined was to have the screen of the laptop cast onto their TV, so that they can continue using the laptop, as is their preference, but on a significantly bigger screen. Although this currently is possible, P8 mentioned significant difficulties in the process of configuring it. One major drawback of VA for P8 was that it felt "overcomplicated" and necessitated a change of habits, requiring them to use the Home Speaker as the central unit, rather than their laptop, which they would strongly prefer. P8 saw potential of voice interfaces for playing music, movies and different radio shows and said that "it should come pre-configured".

\subsubsection{On Familiarity and Accessibility of Design}
Similarly, installing applications on their smartphone was also difficult, even though P8 uses memory hacks to be able to control the smartphone, such as recalling the colors and location of the application icons and navigating by them. This strongly suggests that\textbf{ from the point of view of accessibility, interface updates and re-branding which involve significant changes in color patterns ought to be avoided.}

P8, in line with other research on adoption of technology by older adults, is \textbf{keen on keeping to their patterns of behavior and preferences and instead of finding a completely new solution,} they would like to improve their existing preferred set-up, that is using the laptop as the central unit.  Overall, the aspect of habits, acquired preferences and familiarity with different modes of interaction was a recurring theme for P8, as well as the problems of finding physical items and devices, and setting them up. Upon seeing the possibilities offered by Google Home, P2 remarked "see, you complained that you can't find different devices, and there is no need to find the smart speaker".

\subsubsection{On Making it Easier for the Caretakers} While our previous group of older adults focused extensively on VA applications for people who have physical disabilities, P8 wished for aid in finding items at home and configuring different devices without the help of others. Moreover, they wished to be able to request to play different CDs and Audiobooks they have collected on an outside drive either on their laptop, or their music tower - currently their friends and family assist with this task. As P2 noticed, this is increasingly possible, as we no longer gather our audio libraries on CDs, but rather as files on our smartphones and computers. A prominent concern for P8 was if the introduction of VA into their house environment would not create troubles for the people coming to assist them. 

\subsubsection{On Interactivity and the Phatic Function of Language}
During the free interaction session with the VA, P8 asked the assistant about the time, and received an answer - however, prompted by the moderator - they added that\textbf{ it would be good if the smart speaker offered feedback and answered "I am listening}" as the Home Speaker may be outside of their field of vision, or just not well visible. Moreover, \textbf{the physical design of smart speakers needs to be improved to allow people who still have some sight to know the status of the device. }

\subsubsection{Summary of Key Results}
Overall, during the semi-structured interview the participants raised a few aspects connected to Smart Homes, media control and assistance of caretakers. For example, P8 expressed willingness to use VA to control the window shutters, they were concerned about the cost of such a solution, and reluctant to use it as a substitute of physical activity from fear of not getting enough exercise. For them, there was no need to use it to turn on the lights, as "light switches don't move". The main benefit was its potential to allow them to continue using their laptop as the central hub, to control the music, video and audio-book resources, as the use of VA removes the need to physically find the CD or DVD and insert it in a drive, which is a precise task. Another benefit was the shortened setup of screen-casting from a laptop to a TV - which is now problematic as it involves multiple steps which also require physical precision. 

\section{Conclusions}

In general, older adults participating in our studies wished to be seen as physically active, so they were apprehensive of voice controlled Smart Home solutions. If they admitted to requiring assistance, they wanted to make sure that the IVA solution they used would help, and definitely not burden their caretakers. 
Overall, our participants appreciated VUIs potential and saw many opportunities for its use in their daily life. They also expressed interest in testing this solution at home.

Based on our preliminary results it seems that \textbf{the adoption of this technology for some older adults may depend on the training which will be available within the device itself (a step-by-step tutorial), the ease of setup and inter-connectivity of their home devices the possible integration with their existing preferences regarding the technology solutions they  use at home.}

Furthermore, on the basis of these observations we pose several questions important from the perspective of older adults coping with progressing vision loss:

\begin{enumerate}
\item Can the use of such solutions spread organically? That is, will the older adults with disabilities such as progressing vision loss share their opinion about VUI with other older adults facing similar challenges?
\item Would the introduction of Voice Assistants to people losing their sight be a factor discouraging them from learning new useful skills, such as Braille, and if so, can voice assistant technology become a full substitute for other solutions and skills used by the blind, provided the users can change some of their habits (e.g. use VOD instead of cable) and learn specific commands?
\item Therefore, should voice assistant technology be considered as a new promising primary interface of assistive services for the blind, or should it rather be seen as a supplementary measure?
\end{enumerate}

We believe that to answer these questions it is crucial to study the usability and entry barriers (precise language and commands) of the use of such solutions for the blind and people with progressing vision loss, as well as the extent to which they can supplement or replace other recommended assistive technologies, which may be more expensive. Another crucial point is the degree to which they may help reclaim the independence of people with progressing vision loss, making them less reliant on help of their caretakers or loved ones and thus increasing their empowerment, both in the use of ICT solutions and in their everyday lives.

\section{Acknowledgements}
We would like to thank older adults from our Living Lab, those affiliated with Kobo Association who participated in this study and all transdisciplinary experts involved with the HASE research initiative (Human Aspects in Science and Engineering) of human-computer interaction labs, including XR Lab at Polish-Japanese Academy of Information Technlology (PJAIT), VR Lab at Institute of Psychology, Polish Academy of Sciences, Emotion Cognition Lab at SWPS University and Laboratory of Interactive Technology of the National Information Processing Institute.

\bibliographystyle{splncs04}
\bibliography{bibliography.bib}

\end{document}